\magnification=\magstep1

\hsize=6truein
\vsize=8.5truein
\hfuzz 3.5pt
\null 
\tenrm

\def \um{{\textstyle {1\over2}}}

\def \tm{{\textstyle {3\over2}}}
\def \tq{{\textstyle {3\over4}}}

\def \no{{\textstyle {9\over8}}}

{\centerline
{POSSIBLE INSTABILITY OF THE VACUUM}}
{\centerline
{IN A STRONG MAGNETIC FIELD}}
\vskip 2pc
{\centerline
{ Giorgio CALUCCI\footnote*{E-mail: giorgio@ts.infn.it}}}
\vskip 1pc
{\centerline
{\it Dipartimento di Fisica Teorica dell'Universit\`a di Trieste,
Trieste, I 34014 Italy}}
{\centerline
{\it INFN, Sezione di Trieste, Italy}}
\vskip 2pc
{\centerline {Abstract}}

\midinsert\narrower\narrower
 The possibility that a static magnetic field may decay through production of
 electron positron pairs is studied. The conclusion is that this decay cannot
 happen through production of single pairs, as in the electric case, but only
 through the production of a many-body state, since the mutual magnetic
 interactions of the created pairs play a relevant role.
 \par 
 The investigation is made in view of the proposed existence of huge
 magnetic field strengths around some kind of neutron stars. 
 \endinsert
 \vfill
 \eject
{\bf 1.Introduction}
\vskip 1pc
 The perturbative calculations in QED have been pushed to an exceptional degree
 of refinement as far as their comparison with the
 high precision experiments. The study of the situations which require a
 genuine non-perturbative treatment is less systematic, also in
 this field, however, there are problems that can be considered classical and 
 have gained a high degree of clarification.
 One of these problems concerns the effects of
 intense external field: the decay of the vacuum subject to a strong
 electric field through emission of $e^{-}\,e^{+}$ pairs is well settled through
 the analysis given by Schwinger, starting from the Euler-Heisenberg
 effective Lagrangian[1,2]. If one considers strong magnetic fields there is 
 a strict analogy
 if we admit the existence of monopoles, in that case the problem is only
 quantitative since the monopoles are expected to be very heavy, but for 
 ordinary
 particles the parallel with the electric case is less straightforward. 
 \par
 The
 difference is that we associate a potential energy to the charge in an 
 electric field, but we cannot do the same for an electric charge in a static
 magnetic field. The possible instability of the vacuum in the magnetic
 field, on the other side, seems interesting to be analyzed in view of the
 guess that extremely high magnetic field, of macroscopic extension, are
 realized in nature, around some neutron stars[3]. If the magnetic field is not
 completely static there is certainly a pair production, it can be seen $e.g.$
 treating the problem with the formalism of the adiabatic approximations[4], but
 in this case the rate of production depends on the square of the time
 derivative of the field. Here a qualitative and partially quantitative
 analysis of the possibility that the some kinds of vacuum instability leads to
 pair production even in static conditions is presented.
 \par
 The main idea is stated in this way: any number of pairs in a given magnetic
 field cannot give rise to instabilities unless we take to some extent into 
 account also their mutual interaction: in fact in this last case every pair
 shields partially the magnetic field in which the other particles lie, the
 overall effect could be a decreasing of the field intensity. This effect, when
 the original field is very strong, may decrease the density of magnetic energy 
 and compensate the cost in energy for the creation of the pairs of charged
 particles. If this is true the effect is largely collective and cannot be
 studied particle by particle. A simple energy balance of this collective
 production is presented in section 2 and the reason
 for believing in the instability are given, since at this point we are in
 presence of an $e^{-}\,e^{+}$ multiparticle state the superimposed Coulomb 
 effects are studied in section 3, some conclusions are presented in the last
 section.
 \vskip 1pc
{\bf 2.Main features of the model}
\vskip 1pc
The physical model is described in these terms: 
there is a classical magnetic field $\vec B$ constant in time 	and 
in it
there is a second quantized electron field $\Psi$. The total energy of the
system is 
$${\cal E}_{T}=\um\int B^2 d^3r+{\cal H}\quad {\rm with}\quad
  {\cal H}=\int \{\Psi^{\dagger}[\vec\alpha\cdot (\vec p+e\vec A)+\beta m]
  \Psi \} d^3r\,. \eqno (1)$$
  
  The usual variational principle, for the stationary case, 
  $\delta {\cal E}_{T}=0$
  yields the equation for the magnetic field. This equation is then brought to
  numerical form by putting it among eigenstates of ${\cal H}$. Two known 
  relations are used :
  $$<\sigma|\partial_s {\cal H}|\sigma>=\partial_s <\sigma|{\cal H}|\sigma>=
  \partial_s E_{\sigma}(s)\eqno (2a)$$
  when ${\cal H}$ depends on the parameter $s$, $|\sigma>$is
  one of its eigenstates and $E_{\sigma}$ is the corresponding eigenvalue.
  Since $E$ will be expressed in terms of $\vec B= curl\vec A$, one must
  remember that:
 $${\delta\over {\delta A_i(\vec r)}}=\epsilon_{ikl}\partial_k
   {\delta\over {\delta B_l(\vec r)}}\,.\eqno (2b)$$
 The standard relation $ curl \vec B=\vec J$, yields then
  $$\epsilon_{ikl}\partial_k B_l=-\epsilon_{ikl}\partial_k
  {\delta\over {\delta B_l}} E\quad i.e.\quad
  B_l=-{\delta\over {\delta B_l}} E+\partial_l U\,. \eqno (3)$$
 The functional derivative is zero in the limit $e\to 0$, so $\partial_lU$
 represents the field in absence of vacuum polarization, it will be denoted by
 $B^{(o)}_l$\footnote*{A clear although artificial way of dealing with this kind
 of boundary condition may be found in the formalism of space-dependent coupling
 constants as proposed by Bogoliubov and Shirkov[5].}.
 \par
 Now we specialize to a definite family of field configurations: let the
 magnetic field be uniform $\vec B=B \vec n$ and the volume of quantization of 
 the system be a prism of height $L$, parallel to $\vec n$, and square section
 of side $R$.
 The multiparticle state is obtained in the most trivial way $i.e.$ by filling
 the one-particle states until some Fermi level and leaving completely empty the
 higher one-particle states.
 It turns out natural to consider two independent parameters in order to fix the
  highest populated level
 one for the longitudinal and one for the transverse degrees of freedom; so it
 results (see $e.g.$ [6]):
 
 $$\sum_{\rm levels}={L\over {2\pi}}\int^K_{-K} dk\cdot
 eB {{R^2}\over {2\pi}}\cdot \sum^N_o \sum_{s=\pm 1}$$
 \par
   Looking for a uniform solution $\vec B^{(o)}$ must be a constant vector
  and $-{\delta E/{\delta B_l}}$ is also a constant playing the role of a
  total magnetization $M_l$.
    \par
  The energy coming from ${\cal H}$ has a constant density $\epsilon=E/V\;,\;
  V=R^2 L$. For 
  notational simplicity all the vectors are projected onto the direction
  $\vec n$ and only these components are now used. The total energy density is 
  $$\epsilon_T =\um B^2 +\epsilon =\um (B^{(o)}+M)^2 +\epsilon\,,\eqno (4)$$ 
  and if it can become smaller that the energy density $\um {B^{(o)}}^2$, 
  we may argue 
  that the vacuum polarization leads to an instability of the magnetic vacuum.
  \par
 The single-particle energy levels are the standard relativistic Landau levels
 $$w_{n,s}(k)=\sqrt{m^2+k^2+eB(2n+1+s)}$$
 so the eigenvalue of ${\cal H}$ is
 $$E(K,N)=2\cdot{1\over {4\pi^2}} LR^2 eB \int^K_{-K} dk \Bigl[2\sum^N_{n=1}
 \sqrt{m^2+k^2+2eBn}+\sqrt{m^2+k^2}\Bigr]\,.\eqno (5)$$
 The factor 2 in front of the whole expression arises when the contribution of 
 the positron states is added to the contribution of the electron states.
 In order to perform the discrete sum the approximation used is
 $$\sum^N_{n=1}\sqrt{F+Gn}\approx {2\over{3G}}[F+(N+\um)G]^{3/2},$$
 which is valid up to terms constant in $N$.
 \par
 In the actual situation $F=m^2+k^2\;,\;G=2eB$. The integration over $k$
 may be carried out completely in a straightforward way; the result is
 complicated and not very transparent. Since we are looking for a possible
 instability we are free to choose the trial state and so the range of the 
 parameters $N$ and $K$, a choice that seems promising is 
 $$NeB>>K^2>>eB \quad ,\quad K>>m\,,$$
  because it allows an expansion in decreasing powers of $\sqrt{NeB}.$
  $$
  \epsilon={1\over{3\pi^2}}[2K(2NeB)^{3/2}+K^3\sqrt{2NeB}]+\cdots=
  aB^{3/2}+c\sqrt{B}+ d+\cdots\,. \eqno (6)$$
  The coefficients $a\,,c$ are defined in the expression, $d$ denotes the
  addendum which does not contains the factor $NeB$.
  \par
  This expression is now used to calculate $M$:
  $$M=-\tm a\sqrt{B}-\um c/\sqrt{B}+\cdots\,.\eqno (7)$$
  Then it is inserted into the expression
  for the total energy difference
  $$\Delta=\um B^2+\epsilon-\um [B^{(o)}]^2=\epsilon+MB-\um M^2\,, $$ 
  with the result:
  $$\Delta=-\um aB^{3/2}-\no a^2 B+\um c\sqrt{B}-(\tq ac-d)\,.
  \eqno (8)$$
  The coefficient of the leading power in $B$ is negative and we may therefore
  conclude that there must exist configurations in which the creation of a
  collective state of electrons and positrons has the effect of lowering the
  total energy notwithstanding the cost in energy of the mass and kinetic terms
  for the charged particles.
 \vskip 1pc

 {\bf 3. Effect of the Coulomb interaction}
 \vskip 1pc
  Since we are now considering the creation of a plasma of $e^{+},e^{-}$ we are
  also led to consider the possible effect of the Coulomb interaction. It may be
   estimated perturbatively by inserting the corresponding two-body operator:
  $${\cal V}_c=\um\int\int \Psi^{\dagger}(\vec r)\Psi (\vec r) 
  {{\alpha}\over {|\vec r-\vec s|}}\Psi^{\dagger}(\vec s)\Psi (\vec s) 
  d^3rd^3s \eqno (9)$$
   between the original states.
    This
  interaction is expected to give a negative contribution to the energy and
  since it does not contain $\vec A$ it does not modify the magnetization. 
  The procedure to deal with this problem can be found in
  standard textbooks[7], the actual calculations become, however, very laborious if
  we want to use the
  correct wave functions of the electron, which are essentially 
  harmonic-oscillator functions. A simplified investigation is here presented:
   the
  particle states are simply represented by plane waves, the sums are cut at the 
  same levels as in the previous case, so a maximum longitudinal momentum $K$
  is used and a maximum transverse momentum $P$ is introduced, with the later
  identification $P^2=NeB$.
  In this way it is possible to get a simple estimate of the Coulomb
  effect, in considering the pair interaction
  we obtain the overall cancellation of the diagonal
  terms $(e^{-},e^{-}),(e^{+},e^{+})$
   with $(e^{-},e^{+})$, which in this case is particularly evident, 
  whereas the exchange terms survive and the contributions of the negative and 
  positive 
  charge add up; the result it obviously definite negative and may be 
  expressed as:
  $$E_c=-4\pi V {1\over {(2\pi)}^6}\int \int
  {{\alpha}\over {(\vec p_1-\vec p_2)^2}}d^3 p_1\,d^3 p_2\cdot 2\,, \eqno (10)$$
  the last factor $2$ comes here also from the sum over the two charge states.
  \par
  The integrand depends only on the difference of the momenta, so three
  integrations are easily performed; defining $\rho=K^2/P^2$ the result is
  $$I=\int \int
  {1\over {(\vec p_1-\vec p_2)^2}}d^3 p_1\,d^3 p_2=128 K^2 P^2\int_o^1
   {{(1-u)(1-u)(1-w)}\over {u^2+v^2+\rho w^2}}dudvdw\; $$
   In the limit $\rho \to 0\,i.e.\,P>>K$ the integral develops a logarithmic 
   singularity which gives the leading term:
   $$I\approx 16\pi K^2 P^2 [\ln P^2/K^2 + const]$$
   With the identification $P^2=NeB$ the leading term of the Coulomb
   energy is:
   $$\epsilon_c=-{{\alpha}\over {8\pi^5}}I=
   -{{2\alpha}\over {\pi^4}}NeB K^2\ln (NeB/K^2) \eqno (11)$$
   Comparing it with the results of eq. (6) we see that the Coulomb energy is
   definitively sub-leading. This results gives us confidence that higher order
   perturbative corrections, which may also affect the magnetization,
    will not destroy the main result of the previous chapter. 
\vskip 1pc
{\bf 4.Conclusions}
\vskip 1pc
  The main conclusion we may draw from this particularly simple model is
 that the indications of an instability of the magnetic vacuum are confirmed.
 The source of this instability is to be found in the dependence of the 
 energy density of the electron field on the magnetic field. The leading term 
 grows as $\epsilon_o\propto B^{\sigma}$ and is obviously positive, 
 the diamagnetic term has
 a leading addendum $B(\partial \epsilon_o/\partial B)$, it carries a minus sign,
 and since
 $1<\sigma <2$ this term over-compensates the positive amount of energy required
 for the creation of pairs, the applied magnetic field $B^{(o)}$
is reduced by the pair creations, it results in fact, for the actual case
$\sigma=\tm$
 $B\approx B^{(o)}-\tm a \sqrt{B^{(o)}}+\no a^2$; the effects of the Coulomb 
 interaction are seen only in the sub-leading terms. 
 The requirement of very large magnetic field is essential in order that the
 classification in leading and sub-leading terms be meaningful, in fact looking
 blindly at eq (8) one would get the impression that the total energy
 continuously decreases with increasing $N$, this is clearly impossible, for a
 given $B^{(o)}$ at a certain value of $N$, the resulting field $B$ becomes too
 small for the whole treatment to be correct. The
 determination of the actual values of $B^{(o)}$ for which the process of
 spontaneous pair creation can happen is not possible within the present
 treatment; the main problem in performing a quantitative estimate of the
 process lies in the fact that it proceeds necessarily via a tunnel effect,
 because the creation of a small number of pairs is not enough to lower the
 magnetic energy.
 \par
 In fact the possible existence of the instability of the
magnetic vacuum is strictly related to the mutual interaction of the created
electron pairs, expressed by the fact that they are located in the field 
modified by the existence
of the other pairs. The existence of an instability when only
non interacting pairs are considered is excluded, in a static field, because
 under these conditions, the effective Euler-Heisenberg Lagrangian, which 
 describes the effect of virtual charged spinors in a given external field, 
 never develops an imaginary part, contrary to what happens for the
electrostatic case[1,8]. The situation is also different from what expected in the
spin-one case, where for very intense magnetic fields instabilities are expected
already at the level of single pair creation[9], but the known charged particles
with spin one are much heavier than the electron so there should not 
be sizable interference between the two processes at the foreseen field 
strengths[3] of the order of $10^{10}$T. 
\par
 It must be, finally, noted the all the virtual effects have been ignored, they are likely
 to be important for very large field strengths, because they renormalize the
 electron charge and may also destroy the spherical symmetry of the Coulomb
 interaction[10], also the photon degrees of freedom coupled with
 the electron field may have a role, here only static interactions were 
 analyzed, but all these and possibly other dynamical features are superimposed
 complications, they do not give, however, any indication of being able to 
 destroy the main conclusion that emerges from the simple analysis presented in
 section 2.
   
\vskip 1pc
{\bf Acknowledgments}
\vskip 1pc
This work has been partially supported by the Italian Ministry of the University
and of Scientific and Technological Research by means of the {\it Fondi per la
Ricerca scientifica - Universit\`a di Trieste }.
 \vskip 1pc

{\bf References}
\vskip 1pc
\item{1.}J.Schwinger, Phys. Rev. 82, 664 (1951) and Phys. Rev. 93, 615 (1954).
\item{2.}V.I Ritus: The Lagrangian function of an intense electromagnetic field
         \par
         A.I.Nikishov: The S-matrix of QED with pair creating external field
         in {\it Issues in intense field QED;} ed. V.L. Ginzburg - Nova
         scientia pub. N.Y. 1987.\par
         Qiong-gui Lin J.Phys. G25,17 (1999)
\item{3.}C.Thompson, R.C.Duncan, Astrophys. J. 408, 194 (1993);\par
         H. Hanami, Astrophys. J. 491, 687 (1997);\par
         K. Hurley {\it et al.} , Nature 397, 41 (1999);\par
         M. Feroci $et.\;al.$, Astrophys. J. 515, L9 (1999).
\item{4.}G. Calucci Modern Phys. L. A 14, 1183 (1999).
\item{5.}N.N. Bogoliubov, D.V.Shirkov: Introduction to the theory of quantized
         fields - Ch.VI Interscience Pub. New York 1959.
\item{6.}L.D.Landau, E.M.Lifshits: Quantum mechanics - Ch. XV \P 112
         Pergamon Press Oxford 1977.
\item{7.}L.D.Landau, E.M.Lifshits: Statistical Physics - Ch. VII \P 80
         Pergamon Press Oxford 1980.
\item{8.}J.S.Schwinger Particles, Sources and Fields, vol.2, Ch.4 Addison Wesley
1973.
\item{9.}N.K.Nielsen and P.Olesen N.Phys. B144 (1978), 376;
\item{ }J. Ambjorn and P.Olesen Nucl Phys B315  (1989), 606.
\item{10.}R. Ragazzon, J. Phys. A: Math. Gen 25, 2997 (1992);\par
         G. Calucci, R.Ragazzon J. Phys. A: Math. Gen 27, 2161 (1994).

\vfill
\eject 
\end
\bye